\documentclass[useAMS]{mn2e}
\usepackage{graphicx}
\usepackage{txfonts}
\usepackage{ctable}
\usepackage{threeparttable}

\title[QSO feedback in the early Universe]{Quasar feedback in the early Universe: the case of SDSS J1148+5251}
\author[R. Valiante, R. Schneider, R. Maiolino, S. Salvadori, S. Bianchi]{Rosa Valiante$^{1}$\thanks{E-mail:
valiante@arcetri.astro.it}, Raffaella Schneider$^{1}$, Roberto Maiolino$^{2}$,
Stefania Salvadori$^3$, Simone Bianchi$^4$ \\
$^{1}$INAF - Osservatorio Astronomico di Roma, via di Frascati 33, 00040, Monteporzio Catone, Italy\\
$^{2}$Cavendish Laboratory, University of Cambridge, 19 J.J. Thomson Ave., Cambridge CB3 0HE, UK \\
$^3$Kapteyn Astronomical Institute, Landlaven 12, 9747 AD Groningen, the Netherlands\\
$^4$INAF - Osservatorio Astrofisico di Arcetri, Largo Enrico Fermi 5, 50125, Firenze, Italy}
\begin{document}

\date{Accepted . Received }

\pagerange{\pageref{firstpage}--\pageref{lastpage}} 
\pubyear{2012}

\maketitle

\begin{abstract}
Galaxy-scale gas outflows triggered by active galactic nuclei have been 
proposed as a key physical process to regulate the co-evolution of nuclear 
black holes and their host galaxies.
The recent detection of a massive gas outflow in one of the most distant 
quasar, SDSS J1148+5251 at $z=6.4$, presented by Maiolino et al. (2012) 
strongly supports this idea and suggests that strong quasar feedback is
already at work at very early times. 
In a previous work, Valiante et al. (2011), we have presented a 
hierarchical semi-analytical model, \textsc{GAMETE/QSOdust}, for the formation
and evolution of high-redshift quasars, and we have applied it to 
the quasar SDSS J1148+5251, with the aim of investigating the star formation
history, the nature of the dominant stellar populations and the origin and
properties of the large dust mass observed in the host galaxy. 
A robust prediction of the model is that the evolution of the nuclear black 
hole and of the host galaxy are tightly coupled by quasar feedback in the form 
of strong galaxy-scale winds.
In the present letter, we show that the gas outflow rate predicted by 
\textsc{GAMETE/QSOdust} is in good agreement with the
lower limit of $3500$ M$_\odot/$yr inferred by the observations. 
According to the model, the observed outflow at $z = 6.4$ is dominated by 
quasar feedback, as the outflow rate has already considerably depleted the gas 
content of the host galaxy, leading to a down-turn in the star formation rate 
at $z < 7-8$. 
Hence, we predict that supernova explosions give a negligible 
contribution to the observed winds at $z=6.4$.
\end{abstract}


\section{Introduction}
The energy released by the black hole (BH) accretion process can be powerful 
enough to drive gaseous galactic-scale outflows. Such a negative feedback 
mechanism is expected to play an important role in the formation and evolution 
of quasars (QSOs), by self-regulating the BH growth and eventually quenching 
star formation, therefore affecting the evolution of the physical properties 
of the
host galaxy (e.g. Silk \& Rees 1998; Granato et al. 2004; 
Di Matteo et al. 2005; Springel et al. 2005; Ciotti et al. 2009; 
Hopkins \& Elvis 2010; Zubovas \& King 2012). 
Moreover, negative quasar feedback is also required in theoretical 
models of galaxy evolution to reproduce the space density and properties of old 
passive and massive galaxies, observed in the local Universe up to 
redshift $z \sim 2$ (e.g. Cimatti et al. 2004; Saracco et al. 2005).  

Observational indications of feedback associated with quasar-driven massive 
outflows come from the detection of broad wings of molecular (CO and HCN) 
emission lines (e.g. Feruglio et al. 2010; Alatalo et al. 2011), 
P-Cygni profiles of FIR OH transitions (e.g. Fischer et al. 2010;
Sturm et al. 2011), 
high velocity neutral gas in absorption and high velocity ionized gas 
(e.g. Nesvadba et al. 2010; Rupke \& Veilleux 2011) 
and from the measurement of absorption column densities (Arav et al. 
2002, 2008) in local quasars (but see also Chelouche 2008).
High velocity and broad [O III] emission observed at $z \simeq 2$ (e.g. 
Alexander et al. 2010; 
Harrison et al. 2012) also trace large-scale gas outflows.
Moreover, direct observational evidences of quasar feedback 
quenching star formation at high redshift have recently been presented 
(e.g. Cano-Diaz et al. 2012; 
Trichas et al. 2012).
 
Quasar driven outflows
have also been observed up to $z\sim 6$, however until recently
these were identified only through Broad Absorption Lines in their rest-frame
UV spectrum (Maiolino et al. 2001, 2004), tracing winds in the vicinity of
the accreting black hole and accounting only for a tiny fraction
of the amount of gas in the whole host galaxy. 
 
In a previous work (Valiante et al. 2011, hereafter V11) we have presented a 
semi-analytical model, \textsc{GAMETE/QSOdust},
for the formation and evolution of high redshift quasars. 
This model has enabled us to constrain the 
star formation history (SFH) and the nature of the stellar populations (e.g. 
the stellar initial mass function, IMF) of the well studied high redshift
quasar SDSS J1148+5251 (hereafter J1148) at z=6.4, by linking the 
evolution of the nuclear black hole (BH) with the chemical properties of the 
host galaxy. 
The main observed and inferred physical properties of J1148
are summarized in Table \ref{tab:j1148}. 
\begin{table*}
\caption{Observed and inferred physical properties of the $z=6.4$ QSO J1148
and its host galaxy. 
See Valiante et al. 2011 for details.
}\label{tab:j1148}
\begin{threeparttable}
\begin{tabular}{|l|l|c|l|} \hline 
 Quantity & Description & Value & Reference\\ \hline
  $M_{1450\AA}$ & Continuum rest frame AB magnitude & $-27.82$ & Fan et al. 2003\\
  $M_h$   & Mass of the host dark matter halo & $10^{13}$ M$_\odot$ &  Fan et al. (2004); V11 \\ 
  $M_{BH}$   & Mass of the super massive black hole & $3^{+3.0}_{-1.0}\times 10^9$ M$_\odot$  &  Barth et al. 2003; Willott et al. 2003\\
  $L_{FIR}$  & Far Infrared Luminosity & $(2.2\pm 0.3)$ L$_\odot$    & V11 and references therein \\
  $M_{dyn}$  & Mass of the dynamical gas for a inclination angle $i=65^\circ$ & $(5.5\pm 2.75)\times 10^{10}$ M$_\odot$ &  Walter et al. 2003\\
  $M_{H_2}$  & Mass of the molecular gas from the CO observations & $1.6\times 10^{10}$ M$_\odot$ &  Walter et al. 2003\\
  $M_{star}$   & Stellar mass computed as $M_{dyn}-M_{H_2}$ &$(3.9\pm 2.75)\times 10^{10}$ M$_\odot$ &  Walter et al.2003\\
  $M_{star}$ & Stellar mass from $M_{BH}/M_{star}\sim 0.002$ & $\sim 2.14\times 10^{12}$ M$_\odot$ & Marconi \& Hunt 2003\\
  $SFR$  & Star formation rate & $(3.8 \pm 0.57)\times 10^3$ M$_\odot/$yr & Bertoldi et al. 2003\\
  $Z$      & metallicity from the observations of the narrow line regions in QSOs & $1.32^{+1.57}_{-1.10}$ Z$_\odot$ & Matsuoka et al. 2009\\
  $M_{dust}$ & Mass of dust & $3.4^{+1.38}_{-1.54}\times 10^8$ M$_\odot$ & V11 and references therein\\ 
 \hline 
 $\dot{M}_{outfl}$ & quasar-driven gas outflow rate & $>3500$ M$_\odot$/yr & Maiolino et al. 2012$^a$\\
 \hline 
 \multicolumn{4}{l}{$^a$ The gas outflow rate has been inferred from the 
                      observations of J1148 only after the model presented in 
                      V11 was published (see text)}
\end{tabular}
\end{threeparttable}
\end{table*}
In V11 we have found that the evolutionary scenarios that allow us to  
reproduce the physical properties of the quasar J1148 and of its host galaxy,
such as the mass of the nuclear BH,
the dynamical, gas and dust masses, are tightly constrained. More specifically,
if stars are assumed to form according to a \textit{standard} IMF, 
such as a Larson IMF with a characteristic mass of $m_{ch}=0.35$ M$_\odot$, 
the total stellar mass exceeds the upper limit set by the observed dynamical
mass by a factor of
3-10, depending on the adopted star formation model (see V11 for
further details). Alternatively, the total stellar mass can be accommodated
within
the dynamical mass limit if a \textit{top-heavy} IMF is assumed for all
stellar populations formed in the host galaxy, with a characteristic stellar
mass of $m_{ch}=5$ M$_\odot$.
  
More important, for the purpose of the present investigation, a fundamental 
prediction of the model was that {\it ''a powerful outflow is launched by the quasar 
during the latest $\sim (100-200)$ Myr of the evolution''.} This is a natural
consequence of continuous negative feedback that operates along the merger
history of the host galaxy and it is required to regulate BH growth, star formation 
and to reproduce the observed dynamical and chemical properties in all of the 
evolutionary scenarios presented in V11. 

Recently, 
Maiolino et al. (2012) have detected, for the first time, a massive gas 
outflow, ascribed to strong quasar feedback, through IRAM PdBI observations of 
the [CII] 158$\mu$m transition in the host galaxy of J1148.
They have revealed  broad extended [CII] wings which indicate a large mass 
($7 \times 10^9$ M$_\odot$) of outflowing atomic gas
Assuming a spherical outflow
and that maximum velocity observed in the wings, $\rm v = 1300 km/s$ is the
de-projected velocity of the outflow, they obtain a conservative 
lower limit on the outflow rate of $3500$ M$_\odot/$yr, which represents the
highest outflow rate ever detected (see Maiolino et al. 2012 for further details
on the observations). The associated kinetic power is barely consistent with
the kinetic power that can be produced by a starburst driven wind,
even assuming a star formation rate of $\sim 3000$M$_{\odot}/$yr as inferred
from the strong far-IR thermal emission detected by
(sub-)mm observations of J1148 (Bertoldi et al. 2003; Beelen et al. 2006).
On the other hand, the kinetic energy
is about 0.6\% of the bolometric luminosity of the
quasar, meaning that the observed outflow is more likely to be powered by
the active galactic nucleus (AGN).
 
Here we show that this observational result is a strong confirmation of the
theoretical models presented in V11. In fact, in all of
the evolutionary scenarios
that successfully reproduce the observed properties of J1148, the outflow rate
at $z = 6.4$ is largely dominated by the AGN and it is consistent with the
observed lower limit. 


In what follows, we adopt a Lambda Cold Dark Matter ($\Lambda$CDM) 
cosmology  with $\Omega_m = 0.24$, $\Omega_\Lambda = 0.76$, $\Omega_b = 0.04$, 
and $H_0 = 73$~km/s/Mpc. 
The age of the Universe at a redshift $z = 6.4$ is 900 Myr. 

\section{Model description}
\label{sec:gamete}
Through a binary Monte Carlo algorithm with mass accretion based on the EPS 
theory, we have simulated 50 hierarchical merger histories of a 
$10^{13}\rm{M_\odot}$ dark matter (DM) halo at $z=6.4$.
As discussed in V11, a $[10^{12} - 10^{13}]$ M$_{\odot}$ host DM halo is 
required to match the observed space density of $z \sim 6$ QSOs.
The dependence of the results on the DM halo mass will be discussed in section \ref{sec:concl}.
Using the code \textsc{GAMETE/QSOdust}, we have followed in a self-consistent 
way the build-up of the J1148 central SMBH and of its host galaxy along these
formation paths (V11). 

The nuclear SMBH forms starting from $10^4 h^{-1}$ M$_\odot$ seed BHs hosted in 
very high redshifts halos. Along each merger tree, the assembly of the host 
galaxy and its SFH are driven by binary mergers and mass accretion. 
Similarly, seed BHs grow by coalescences with other BHs and by gas accretion.
The co-evolution of these two components, the SMBH and its host galaxy, is
mostly controlled by quasar feedback in the form of a galactic-scale wind, 
which self-regulate the BH growth and eventually halts star formation.

The evolution of the mass of gas, stars and metals in the 
interstellar medium (ISM) is followed by taking into account infall and outflow 
processes from/to the external medium.
Metal and dust enrichment is implemented in all progenitor galaxies by 
considering the contribution from stars of different masses, such as 
Asymptotic Giant Branch (AGB) stars and Supernovae (SNe), taking into account 
the proper stellar lifetimes. Moreover, 
subsequent dust grain re-processing in the ISM, namely destruction by 
interstellar shocks and grain growth in molecular clouds (MCs), is also taken 
into account.

Finally, different SFHs for the QSO host galaxy are explored: 
\textit{quiescent} models, where the efficiency of star formation is 
independent of galaxy mergers, and \textit{bursted} models 
in which the star formation efficiency is enhanced during major mergers between
progenitor galaxies.

\section{Model Results}
\label{sec:j1148}
In this section we present the models that in V11 have 
been identified to reproduce the available observational constraints on J1148.  
In all models, the free parameters, namely the efficiency of star formation, 
BH accretion and quasar-driven wind have been chosen to reproduce the BH and 
gas mass of J1148. 

As discussed in V11, the inferred stellar mass, a crucial information for the 
evolutionary models, is affected by large uncertainties. This is
the reason why two different values are given in Table \ref{tab:j1148} 
and the models that we have explored 
adopt progressively higher star formation (SF) efficiencies, and therefore 
larger final stellar masses. We thus, investigate the corresponding effects 
on the evolution of the chemical properties of host galaxy (metal and dust 
content). 

Four out of the several models presented in V11 have been found to 
successfully reproduce the properties of the QSO J1148: 
\textbf{Q1 t.h.}, \textbf{B1 t.h.}, \textbf{Q2} and \textbf{B3}. The
first two are quiescent (Q1 t.h.) and bursted (B1 t.h.) star
formation models where stars form according to a top-heavy IMF, while  
the quiescent (Q2) and bursted (B3) models assume a 
standard IMF 

\subsection{BH-M$_{star}$ relation}\label{sec:bhrel}
\begin{figure}
\centering
\includegraphics [width=8.0cm]{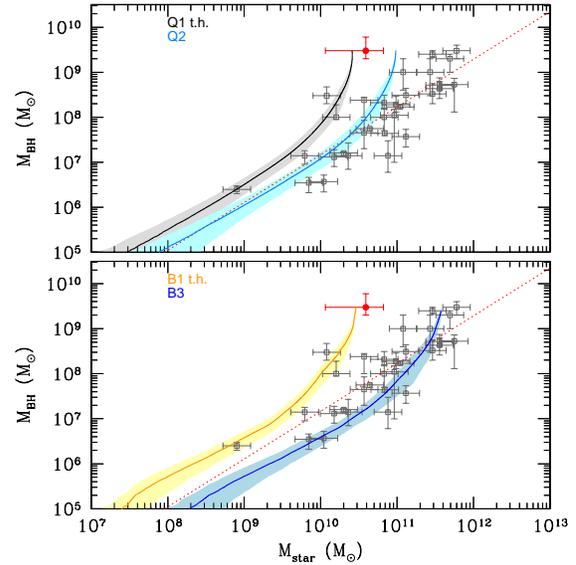}
\caption{The evolution of the J1148 BH mass as a function of the stellar mass 
         for quiescent (upper panels, Q1 t.h. with black
         line and Q2 with azure line) and bursted models (lower panel, B1 t.h.
         gold line and B3 blue line).
         The filled circle and open squares are the J1148 and local QSOs and 
         galaxies observed M$_{BH}-$M$_{star}$ relations, respectively. 
         The dashed line is the Marconi \& Hunt empirical fit to the local 
         relation. Solid lines are the four different models that in V11 have 
         been selected to reproduce the available observations of J1148 (see 
         text), averaged over 50 different merger tree realizations. The shaded
         areas represent the 1$\sigma$ dispersion.} 
\label{fig:bhrel} 
\end{figure}

For convenience, we report in figure \ref{fig:bhrel} the evolution of the 
M$_{BH}-$M$_{star}$ relation predicted by the four models introduced above.
The location of J1148 on the M$_{BH}-$M$_{star}$ plane 
is indicated by the solid circle (with M$_{star}$ given by the observations, see 
Table \ref{tab:j1148}) and is compared with the M$_{BH}-$M$_{star}$ 
relation observed in local quasars and galaxies (open squares) and the 
empirical fit taken from Marconi \& Hunt (2003). 
A detailed discussion can be found in V11. 
By construction, all models reproduce the mass of the SMBH in J1148 at z=6.4,
but predict different final stellar masses 
going from the value inferred from the observations, $\sim 3.9\times 10^{10}$ 
M$_\odot$ (models Q1 t.h. and B1 t.h.) up to values progressively closer to 
the observed local relation, $\sim (1-4)\times 10^{11}$ M$_\odot$ (Q2 and B3, 
respectively).  

\subsection{Chemical properties and SFH}\label{sec:chemevo}
\begin{figure}
\centering
\includegraphics [width=8.0cm]{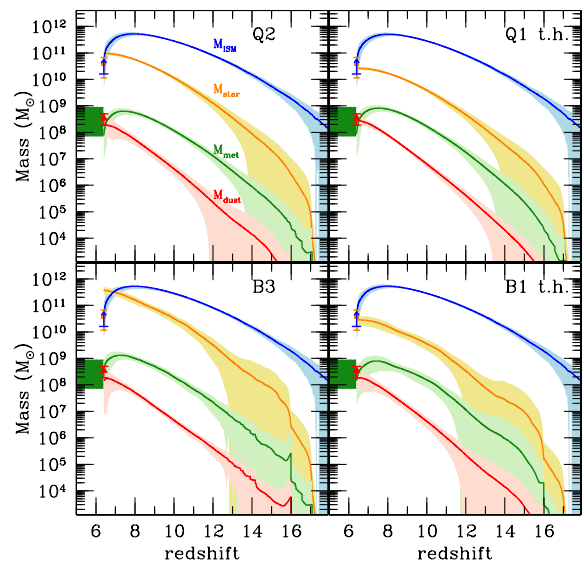}
\caption{The ISM chemical evolution of the J1148 host galaxy in the four 
         different models which best reproduce the observed properties. 
         Solid lines are the evolution of the mass of gas (blue),
         stars (yellow), metals (green) and dust (red) averaged over 50 
         hierarchical merger histories with the 1$\sigma$ dispersion given by 
         the corresponding shaded areas. The blue arrow indicates the 
         molecular gas mass ($M_{H_2}$), considered as a lower limit for the 
         total gas mass, the yellow solid square is the stellar mass computed 
         as the difference between the dynamical and molecular gas mass, the 
         red triangle is the mass of dust and the green region represents the 
         mass of metals ($Z\times M_{H_2}$).} 
\label{fig:chemevo} 
\end{figure}

By using the chemical evolutionary model with dust described in V11, we have 
analyzed the dependence of the predicted J1148 chemical properties 
(in particular of metals and dust) on both the SFH (quiescent vs 
bursted star formation rates, SFRs) and IMF (standard vs top-heavy).
The results of the four selected models are shown in figure
\ref{fig:chemevo}. In this plot, we present the predicted evolution of the mass 
of gas, stars, metals and dust compared with the available observations for 
J1148. 
As it can be seen from the figure, independently
of the SFH and IMF, the mass of gas and metals both grow up to a maximum value, 
reached at redshift $z\sim 8$, \textit{and than decrease due to the 
strong effect of the quasar feedback}. 

The SFR of high redshift quasars is usually estimated from the FIR luminosity,
adopting the $L_{FIR}-\rm{SFR}$ relation derived by Kennicutt (1998) under
the assumption that the dominant dust-heating mechanism is radiation from 
young stars.
The conversion factor usually adopted to convert $L_{FIR}$ to SFR,
$\sim 5.8\times 10^9$ L$_\odot$, has been derived assuming a $10-100$ Myr old
burst of star formation and a Salpeter IMF (see Kennicutt 1998).
Following this prescription, the FIR luminosity estimated for J1148 
(see Table \ref{tab:j1148})
gives a SFR $\sim (3.8\pm 0.57)\times 10^3$ M$_\odot/$yr (Bertoldi et al. 2003). 
However, it has been pointed out
that the inferred SFR would be about an order of magnitude lower 
($\sim 180$ M$_\odot/$yr) if the Schmidt-Kennicutt law is instead adopted 
(Dwek et al 2007; Li et al. 2008; V11).

In this work we have further revised the SFR inferred from the observations, 
by taking into account the different IMFs adopted in our models and 
we rescale the $L_{FIR}-\rm{SFR}$ conversion factor for a Larson IMF with
$m_{ch}= 0.35$ M$_\odot$ (standard IMF models) and $5$ M$_\odot$ (top-heavy IMF). 
With this correction we obtain star formation rates
$\sim 3$ to $30$ times lower, more specifically: 
SFR $\sim (1087\pm 163) $M$_\odot/$yr, for a standard
IMF, 
and SFR $\sim (113\pm 17)$ M$_\odot$/yr, for a top-heavy IMF. 
These values are still consistent with results quoted in 
previous works (Maiolino et al. 2005; Dwek et al. 2007; Li et al. 2008).
Yet, as discussed in V11, we use these values as an upper limit to the 
effective rate of star formation in J1148, since the active quasar itself may 
give a non negligible contribution to dust heating 
(Bianchi, Valiante \& Schneider in preparation).
The quiescent and bursted SFHs 
for the four selected models are shown in the left panels of
figures \ref{fig:out_std} and \ref{fig:out_th}. 

\section{Quasar and SN-driven gas outflow}\label{sec:outflow}
In \textsc{GAMETE/QSOdust}, galactic outflows can be driven by 
SN explosions and by the AGN. 
Both stellar and quasar feedback are consistently modelled in the form of 
energy-driven winds sweeping the surrounding material away from the galaxy. 
The total gas outflow rate is given by $dM_{ej}(t)/dt = dM_{ej,SN}(t)/dt + 
dM_{ej,AGN}(t)/dt$, with:
\begin{equation}
\frac{dM_{\rm ej,AGN}}{dt} = 2 \epsilon_{\rm w,\textsc{agn}} \epsilon_{\rm r} \big(\frac{c}{v_{\rm e}} \big)^{2} \dot{M}_{\rm accr}  
\label{eq:agnfdb}
\end{equation} 
\noindent 
and
\begin{equation}
     \frac{dM_{\rm ej,SN}}{dt}=\frac{2 \epsilon_w 
       E_{\rm SN}}{v_{\rm e}^2} R_{\rm SN}(t),
     \label{eq:snfdb}
\end{equation}
where  $v^2_{\rm e}$ is the escape velocity of the galaxy, 
$R_{\rm SN}(t)$ is the SN explosion rate, 
$\dot{M}_{\rm accr} $ is the BH gas accretion rate, 
$\epsilon_{\rm r}$ is the radiative efficiency, fixed to be 0.1 
and $E_{\rm SN}$ is the average supernova explosion energy, assumed to be 
$2.7\times 10^{52}$ erg for PISNe and $1.2\times 10^{51}$ erg for core collapse 
SNe.
The efficiency of SN-winds ($\epsilon_w=2\times 10^{-3}$) has been 
calibrated by the chemical evolution model applied to the Milky Way and its 
dwarf satellites (Salvadori et al. 2007, 2008).
The AGN-wind efficiency ($\epsilon_{w,AGN}=5\times 10^{-3}$) and the BH accretion
efficiency ($\alpha=[180-200]$, according to the particular model), 
have been instead chosen to reproduce at the same time the gas and BH mass of 
J1148.

\begin{figure}
\centering
\includegraphics [width=8.0cm]{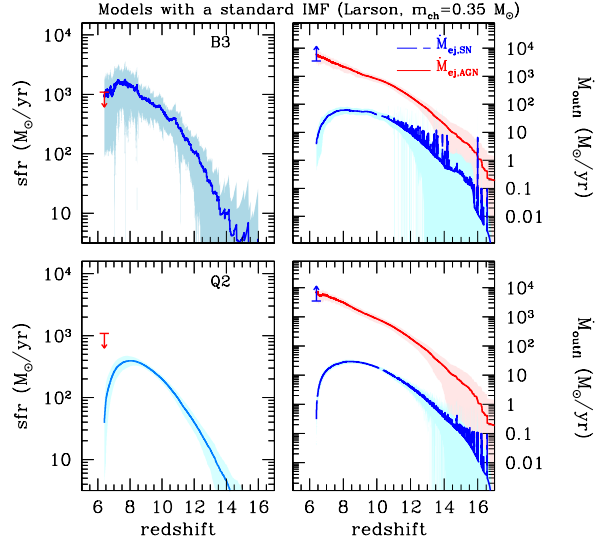}
\caption{The star formation (left panels) and the gas outflow (right panels) 
         rates as a function of redshift, predicted by models B3 and 
         Q2. In both models a Larson IMF with a characteristic mass 
         $m_{ch}=0.35$ M$_\odot$ has been adopted. All the curves are the 
         averages over 50 random merger tree realizations of the quasar host 
         galaxy, with shaded areas representing the 1$\sigma$ dispersion.
         Blue dashed and red solid lines in right panels represent the 
         mass of gas ejected per unit time by SN and quasar driven winds, 
         respectively. The arrows indicate the upper limit to the SFR of 
         J1148, corrected for our adopted IMFs or to the gas 
         outflow rate, obtained by Maiolino et al. (2012).} 
\label{fig:out_std} 
\end{figure}

\begin{figure}
\centering
\includegraphics [width=8.0cm]{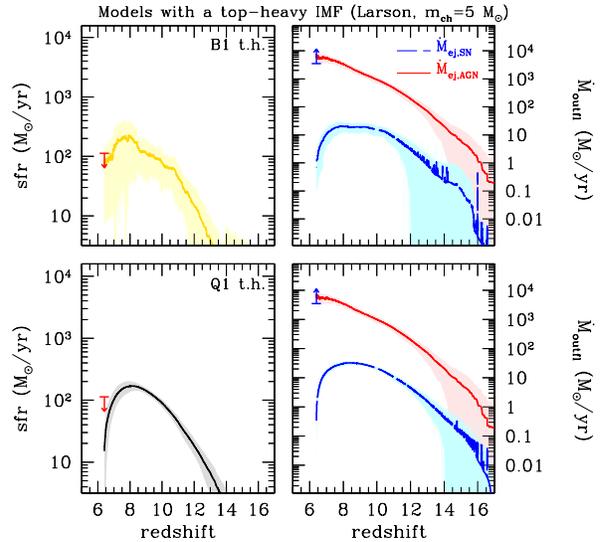}
\caption{The same as in figure \ref{fig:out_std} but for models B1 and 
         Q1. In these two models a Larson IMF with $m_{ch}=5$ M$_\odot$ has been
         adopted.} 
\label{fig:out_th} 
\end{figure}

The right panels of figures \ref{fig:out_std} and \ref{fig:out_th} show the
redshift evolution of the gas outflow rate predicted by the selected models, 
as labelled in the figures, by showing the separate contribution of SNe and 
the quasar.
As it can be seen from these figures, for $z<12$ the QSO outflow rate
exceeds by more than two orders of magnitude the SN outflow rate in all the 
different models.

For all models, the predicted outflow rate at $z=6.4$ is fully consistent with 
the lower limit to the gas outflow rate inferred by Maiolino et al. (2012).
With this large outflow rate, the host galaxy should be rapidly cleared of 
its gas content and star formation
should be quenched. Indeed, the predicted SFRs systematically 
show a down-turn at redshifts $z<7-8$, when the 
effect of the quasar-driven wind on the evolution of the ISM gas mass becomes 
dramatic (see blue lines in figures \ref{fig:chemevo}).  
Such a decline is reflected also into the SN rate and, therefore, in the 
SN-driven outflow rate, which is orders of magnitude lower than the 
quasar-driven one.

The formulation adopted in V11 for the BH accretion rate ensures that
the quasar feedback is active over almost the entire lifetime of the host halo.
This feedback results into efficient gas ejection at $z<8$, when
the BH mass and accretion rate are higher (see V11 for a detailed discussion).
 
\section{Conclusions}\label{sec:concl}

In this work, we have shown that the formation history of SMBH at the center
of high redshift quasars is strongly linked to the properties of their host 
galaxies.

We use a state-of-the-art model (V11) that allows us to simulate a large 
number of independent hierarchical histories of quasar host galaxies, 
following at the same time the star formation histories, chemical evolution, 
and the gradual build-up of the nuclear black hole by mergers and gas 
accretion. 

We have recently applied this numerical model to one of the most distant 
quasars, SDSS J1148 at $z=6.4$ and we have shown that the joint constraints 
provided by the dynamical, gas, and dust masses together with the mass of the 
nuclear SMBH, allow to reconstruct the star formation histories and the nature 
of the dominant stellar populations in these systems. 
{\it A robust prediction of these models
 is that the evolution of the nuclear black hole and of the host galaxy are 
tightly coupled by quasar feedback in the form of strong galaxy-scale winds.}
 
The recent detection of broad wings in the [CII]158$\mu$m line associated to
SDSS J1148 by Maiolino et al. (2012), is a strong evidence of clear quasar 
feedback in the early Universe. 
 
The comparison between model predictions and observational data shows that the
gas outflow rates predicted by the models are fully consistent with the lower
limit ($3500$ M$_\odot/$yr) inferred from the observations. 
In addition, it lends strong support to the interpretation that the dominant
driving mechanism is quasar feedback. 
According to the physical prescriptions and parameter values adopted 
in the models presented here,
the outflow rate has already significantly 
affected the gas content of the galaxy, leading to a down-turn in the star 
formation rate at $z < 7-8$. 

It is important to stress that the theoretical findings have been derived
independently, well
before the observations. Therefore, strong quasar feedback has been a 
natural by-product of models that have been identified to reproduce the 
observed dynamical and chemical properties of the host galaxy.  
In all models, the SN rate is constrained by the SFR inferred by the 
observations and by the chemical properties of the host galaxy. The 
associated SN-wind efficiency has been calibrated using the same chemical
evolution model to reproduce the global properties of the Milky Way Galaxies
(Salvadori et al. 2008).
To be at least comparable with the observed outflow rate, the SN wind efficiency
should be higher by
$\sim 3-4$ orders of magnitude.
There are no physical motivations to justify such an enormous efficiency.

All the proposed scenarios, i.e. different combinations of SFHs and IMFs, 
provide consistent results for the quasar-driven gas ejection rate. 
Thus, the recent observation reported by Maiolino et al. (2012) does not 
help to break the degeneracy between the proposed models: accurate observations 
of the stellar mass are necessary to further constrain the properties of the 
J1148 host galaxy therefore helping in the identification of 
the most appropriate IMF and SFH.
  
Finally, the comparison of our predicted gas outflow rate with that observed 
for J1148 by Maiolino et al. (2012) allow us to put additional constraints on 
the mass of the DM halo hosting the SMBH. 
With a DM halo mass equal to $\rm 10^{12} M_\odot$, both the BH 
accretion and feedback efficiencies need to be reduced in order to reproduce 
the observed BH and gas masses at $z = 6.4$. In fact, a smaller DM halo mass 
implies a reduced gas mass available at each redshift along the hierarchical
history. As a result, the final outflow rate is more than one order of 
magnitude smaller than for the reference model with DM halo mass of 
$\rm 10^{13} M_\odot$ and inconsistent with the observed lower limit.

\section*{Acknowledgments}
We would like to thank the anonymous Referee for useful comments and 
suggestions. This work has been supported by the program PRIN-INAF 2010 
through the grant $"$The 1 billion Year Universe: Probing Primordial Galaxies 
and the Intergalactic Medium at the edge of Reionization$"$.


\label{lastpage}

\end{document}